\documentclass[journal, a4paper]{IEEEtran}

\IEEEoverridecommandlockouts

\usepackage{graphicx}
\usepackage[subrefformat=parens]{subcaption}
\usepackage{subfloat}
\usepackage{float}
\usepackage{amsmath,amsthm,amsfonts,amssymb}
\usepackage{cite}
\usepackage{bm}
\usepackage{bbm}
\usepackage{url}
\usepackage{array}
\usepackage{color}
\usepackage{multirow}
\usepackage{booktabs,multirow}
\usepackage[table,xcdraw]{xcolor}
\usepackage{enumerate}
\usepackage{enumitem}
\usepackage{makecell}
\usepackage{comment}
\usepackage{caption}

\usepackage[english]{babel}
\usepackage{algorithm}
\usepackage[noend]{algpseudocode}

\theoremstyle{plain}

\usepackage{mathtools}

\usepackage[english]{babel}
\usepackage{algorithm}
\usepackage[noend]{algpseudocode}

\allowdisplaybreaks[4]

\begin{document}
\title{DeepJSCC-l++: Robust and Bandwidth-Adaptive Wireless Image Transmission
}

\author{
Chenghong~Bian,
Yulin~Shao,~\IEEEmembership{Member,~IEEE},
Deniz~G\"und\"uz,~\IEEEmembership{Fellow,~IEEE}
\thanks{C. Bian and D. G\"und\"uz are with the Department of Electrical and Electronic Engineering, Imperial College London, London SW7 2AZ, U.K. (e-mail: \{c.bian22, d.gunduz\}@imperial.ac.uk). Y. Shao is with the State Key Laboratory of Internet of Things for Smart City, University of Macau, Macau S.A.R. (e-mail: ylshao@um.edu.mo).
}
\thanks{This work  was supported in part by UKRI for the project AIR (ERC-CoG, EP/X030806/1), and in part by the IOTSC-UM Conference Grant SKL-IoTSC(UM)-2021-2023 and the start-up grant SRG2023-00038-IOTSC.}
}

\maketitle

\begin{abstract}
This paper presents a novel vision transformer (ViT) based deep joint source channel coding (DeepJSCC) scheme, dubbed DeepJSCC-l++,  which can adapt to different target bandwidth ratios as well as channel signal-to-noise ratios (SNRs) using a single model. To achieve this, we treat the bandwidth ratio and the SNR as channel state information available to the encoder and decoder, which are fed to the model as side information, and train the proposed DeepJSCC-l++ model with different bandwidth ratios and SNRs. The reconstruction losses corresponding to different bandwidth ratios are calculated, and a novel training methodology, which dynamically assigns different weights to the losses of different bandwidth ratios according to their individual reconstruction qualities, is introduced. Shifted window (Swin) transformer is adopted as the backbone for our DeepJSCC-l++ model, and it is shown through extensive simulations that the proposed DeepJSCC-l++ can adapt to different bandwidth ratios and channel SNRs with marginal performance loss compared to the separately trained models. We also observe the proposed schemes can outperform the digital baseline, which concatenates the BPG compression with capacity-achieving channel code. We believe this is an important step towards the implementation of DeepJSCC in practice as a single pre-trained model is sufficient to serve the user in a wide range of channel conditions. 

\end{abstract}


\section{Introduction}\label{sec:intro}

Thanks to recent advances in machine learning, there has been a growing interest in developing data-driven joint source-channel coding (JSCC) systems. Focusing on the wireless transmission of images, the DeepJSCC scheme proposed in \cite{JSCC2019} is shown to achieve better performance and enhanced robustness against channel variations compared to conventional separation-based baselines. Over the last several years, DeepJSCC approach has been successfully extended to many new scenarios, exhibiting its potential as a viable technology, from multi-path fading channel \cite{jsccofdm, Wu:WCL:22} to multi-input multi-output (MIMO) \cite{vit_mimo,jsccmimo} and multi-user scenarios \cite{jsccrelay}.


On the other hand, in most existing works, the DeepJSCC encoder/decoder pairs are designed and trained for specific channel conditions, i.e., channel bandwidth and signal-to-noise ratio (SNR). This is a limitation for the adoption of DeepJSCC in practical systems, as it requires storing a large number of DeepJSCC encoder/decoder parameters on mobile devices to be used in different channel conditions, imposing significant memory requirements. In \cite{xu2021wireless, Wu:WCL:22}, it has been shown that a single DeepJSCC network can adapt to different channel SNRs. In this work, we will show that a single DeepJSCC encoder/decoder pair can be trained to be used in any available channel bandwidth and SNR. This not only shows the flexibility of DeepJSCC, but significantly increases its potential to be applied in practice. 

DeepJSCC with a varying bandwidth is also studied in \cite{deepjsccl}; however, \cite{deepjsccl} considers a \textit{successive refinement} scenario, where transmission takes place over several channel blocks, and the image should be recovered from any number of first $l$ channel blocks (Fig. \ref{fig:system_model:a}). This scenario is applicable when broadcasting an image to several receivers, where each receiver can receive a different number of channel blocks. Instead, we consider a single receiver, where the available channel bandwidth can vary from image to image, but is known for each image prior to transmission, dictated, for example, by the latency constraint of the underlying application. Alternatively, variable-length transmission is studied in \cite{bw_adapt, predictive_adapt, bupt}, where the encoder decides how much channel bandwidth to use based on the content of the input image. 



In this paper, we propose a novel bandwidth and channel quality adaptive scheme, named DeepJSCC-l++, which can map each input image to a desired channel bandwidth - see Fig.  \ref{fig:system_model:b}. The DeepJSCC-l++ encoder takes the image as well as the bandwidth ratio and the channel SNR as side information to produce the codeword. We introduce a novel code architecture using the Swin transformer \cite{swin} as the backbone. To balance the reconstruction qualities for different bandwidth ratios, a novel training methodology, called \textit{dynamic weight assignment (DWA)}, is introduced, which assigns different weights to the losses corresponding to different bandwidth ratios. Simulation results show the effectiveness of the proposed DeepJSCC-l++ scheme, which can be adaptive to both the bandwidth ratio and the channel SNR with negligible gap from the reconstruction performance obtained with separately trained models. This shows that the proposed architecture is capable of acquiring and prioritising the input image features, and sending only the most important features depending on the available bandwidth, while employing the necessary amount of redundancy against channel noise depending on the channel SNR. 

We highlight that the proposed DeepJSCC-l++ architecture can also be applied  to the successive refinement problem. Our results show that DeepJSCC-l++ provides a significant performance improvement compared to the CNN-based solution in \cite{deepjsccl} in this scenario. Finally, DeepJSCC-l++ outperforms the separation-based baseline that combines BPG compression algorithm with a capacity-achieving channel code. This result shows the potential of the transformer architecture in designing state-of-the-art DeepJSCC solutions. 

\section{System Model}\label{sec:II}
We consider the wireless transmission of images over the AWGN channel. Let $\bm{S} \in \mathbb{R}^{C\times H\times W}$ denote the input image, where $C, H, W$ denote the number of color channels, height, and width of the image, respectively. We define $N = CHW$ to denote the input dimension to facilitate following definitions. The encoder maps the input image, $\bm{S}$, to a complex codeword $\bm{z} \in \mathbb{C}^{\rho N}$, where $\rho N \in \mathbb{Z}$, denotes the bandwidth used for transmission. Here, $\rho$ is defined as the bandwidth ratio as it represents the average number of channel symbols available per source dimension. The transmitted codeword $\bm{z}$ goes through an AWGN channel, $\bm{y} = \bm{z} + \bm{w}$, where each element in $\bm{w}$ follows a complex Gaussian distribution with zero mean and variance equals to $\sigma^2$, and $\bm{y}\in \mathbb{C}^{\rho N}$ denotes the noisy channel output vector. We impose an input power constraint: $\frac{1}{\rho N}\|\bm{z}\|_2^2\leq 1$ for each $\bm{z}$, which means that the channel SNR is given by $\mathrm{SNR} = 1/\sigma^2$. 

The decoder maps the received vector directly to its estimate of the input signal, $\widetilde{\bm{S}} \in \mathbb{R}^{C\times H\times W}$. The reconstruction quality can be measured through a variety of distortion measures. In this paper, we will consider the most commonly used peak signal-to-noise ratio (PSNR), defined as:
\begin{align}
    \text{PSNR} &= 10\log_{10} \frac{255^2}{\frac{1}{N}||\bm{S}-\widetilde{\bm{S}}||^2_F}.
    \label{eq:psnr}
\end{align}
glo
In conventional separate source-channel coding schemes, we choose a pair of compression and channel coding rates depending on the channel SNR and bandwidth ratio, $\rho$. Channel coding rate depends on the channel SNR, and is chosen to guarantee reliable transmission with high probability. The source compression rate is dictated by the channel code rate and the available bandwidth ratio, $\rho$. In practical systems, a mobile device chooses from a list of prescribed list of modulation and coding schemes (MCSs) according to the estimated channel SNR. In the case of JSCC, since we have a single code, the code parameters will depend on both the SNR and the bandwidth ratio. The initial works on DeepJSCC \cite{JSCC2019} considered a similar approach to these practical systems, where a separate DeepJSCC encoder/decoder pair is trained for given $(\mathrm{SNR}, \rho)$ pairs. However, due to the high memory complexity of DeepJSCC codes, it is not practically feasible to assume that each device can store a large variety of code parameters to be used in different channel conditions. Therefore, the goal in this paper is to train a single encoder/decoder pair, which can dynamically adapt to the desired bandwidth ratio and $\mathrm{SNR}$ level in an online manner at the time of transmission. 

Let $f_{\Theta}$ denotes the encoder function parameterized by DNN parameters $\Theta$, while the decoder is denoted by $g_{\Phi}$, paramaterized by $\Phi$. A single DNN model will be utilized to adapt to different $\mathrm{SNR}$ and $\rho$ values, where the encoder $f_{\Theta}$ takes the image $\bm{S}$ as well as the $\mathrm{SNR}$ and $\rho$ as input, and we have $\bm{z} = f_{\Theta}(\bm{S}, \mathrm{SNR}, \rho) \in \mathbb{C}^{\rho N}$. The receiver then reconstructs the original image as $\widetilde{\bm{S}} = g_{\Phi}(\bm{y}, \mathrm{SNR}, \rho) \in \mathbb{R}^{C\times H\times W}$. For the sake of simplicity, we will assume that only $L$ different bandwidth ratios are allowed, specified by $\{\rho_1, \ldots, \rho_L\}$, where we assume that $\rho_l = l \cdot \rho_1$, $l \in [L]$.


\begin{figure}[t]
     \centering
     \begin{subfigure}{\columnwidth}
         \centering
         \includegraphics[width=0.7\columnwidth]{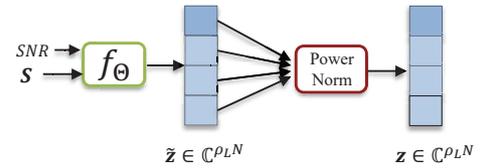}
         \caption{Successive refinement: Each image is transmitted using the full bandwidth, but can be recovered from any of the first $l$ channel blocks.}
         \label{fig:system_model:a}
     \end{subfigure}     
     \begin{subfigure}{\columnwidth}
         \centering
         \includegraphics[width=0.7\columnwidth]{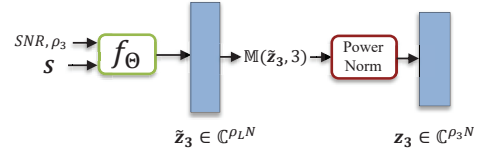}
         \caption{Bandwidth-adaptive transmission: Available channel bandwidth is dictated by the application, and may change from image to image.}
         \label{fig:system_model:b}
     \end{subfigure}

  \caption{Successive refinement vs. bandwidth-adaptive transmission.}
  \label{fig:system_model}
\end{figure}

\begin{figure*}[t]
  \centering
  \includegraphics[width=0.7\linewidth]{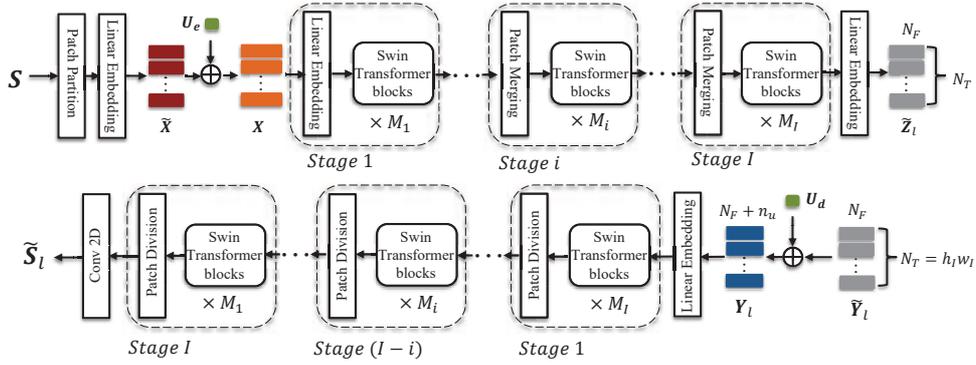}\\
  \caption{The detailed neural network architecture for the proposed DeepJSCC-l++ model.}
\label{fig:detailed_net}
\end{figure*}

Next, we present the successive refinement problem formulation considered in \cite{deepjsccl} and highlight the differences with respect to the adaptive-bandwidth formulation considered here. As shown in Fig. \ref{fig:system_model:a}, in the successive refinement scheme studied in \cite{deepjsccl}, the encoder maps the image into a latent vector $\tilde{\bm{z}} \in \mathbb{C}^{\rho_LN}$, which is further power normalized to $\bm{z}$ before transmission. Then, $L$ different receivers are considered, where receiver $l$ receives only the first $l$ portions of the noisy signal. That is, if we define $\bm{y}^\top = \bm{y}^\top_{1:L} =  [\bm{y}_1^\top, \cdots, \bm{y}_L^\top]$, then the receiver $l$ reconstructs the signal as $\widetilde{\bm{S}}_l = g_{\Phi, l}(\bm{y}_{1:l}, \mathrm{SNR})$. Therefore, in this scheme, the goal of the transmitter is to generate a codeword that can simultaneously satisfy $L$ receivers, each of which receives a different amount of information. Accordingly, the additional bandwidth available to receiver $l$ is used to refine the reconstruction generated by receiver $(l-1)$. Additionally, the formulation in \cite{deepjsccl} imposes a separate power normalization to each part of the transmitted codeword.

We note that the DeepJSCC-l scheme proposed in \cite{deepjsccl} with successive refinement is also a solution to our problem, although we expect it to be a suboptimal solution since in our problem we have a more relaxed requirement: the transmitter knows the available bandwidth, and can adapt its transmission accordingly, as a consequence, a single average power allocation can be used.

\subsection{Proposed solution: DeepJSCC-l++}

Here, we propose a new architecture called DeepJSCC-l++. In the proposed solution, the encoder maps the input image to an output of dimension $\rho_L N$, denoted by $\tilde{\bm{z}}$, similarly to DeepJSCC-l. However, instead of transmitting the full-bandwidth codeword, we adopt a simple mask at the encoder to transmit only the first $\rho_l N$ elements, denoted by $\bm{z}_l$, which is expressed as
\begin{align}
    \bm{z}_l = \mathbb{M}(\tilde{\bm{z}}, l),
    \label{eq:masking}
\end{align}
where $\mathbb{M}(\cdot, \cdot)$ represents the masking operation. Note that $\bm{z}_l$ is subject to a more flexible power constraint compared with the successive refinement scheme: $\frac{1}{\rho_l N}{||\bm{z}_l||^2_2} \le 1$. The power normalized signal is then transmitted over the complex AWGN channel. At the receiver, the decoder zero-pads the received signal $\bm{y}_l$ to a length-$\rho_L N$ vector, $\tilde{\bm{y}}$, and takes both $\tilde{\bm{y}}$ and the side information $\rho_l, \mathrm{SNR}$ as input to reconstruct $\widetilde{\bm{S}}_l = g_{\Phi}(\tilde{\bm{y}}, \mathrm{SNR}, l) \in \mathbb{R}^{C\times H\times W}$ using a decode function $g_{\Phi}: \mathbb{R}^{\rho_L N}\times \mathbb{R} \times \mathbb{Z} \rightarrow \mathbb{R}^{C\times H \times W}$.


\section{Methodology}\label{sec:III}
In this section, we present the neural network architectures to parameterize the encoder and decoder $f_{\Theta}$ and $g_{\Phi}$ for the DeepJSCC-l++. Then we propose a novel training methodology called DWA to avoid substantial performance loss at the higher bandwidth ratios.

\subsection{Neural Network Architectures}\label{sec:IIIA}
As shown in Fig. \ref{fig:detailed_net}, we use one of the state-of-the-art vision transformer models, the Swin transformer \cite{swin}, as the backbone, and the data flow for the proposed scheme is detailed as follows: 

\textbf{Data flow at the encoder.} We start with the initial stage where  the image $\bm{S}$ is first split into non-overlapping patches (also known as `tokens') by a patch partition module followed by a linear embedding layer to project each `token' into feature space with dimension $c$. We use a patch size of $2 \times 2$ thus the aforementioned patch partition and linear embedding modules transform the original image from dimension $C\times H\times W$ to a feature tensor $\widetilde{\bm{X}}$ with dimension $c\times H/2\times W/2$. Before feeding $\widetilde{\bm{X}}$ into the subsequent transformer layers, we concatenate each of its tokens with the side information, $\bm{u} \in \mathbb{R}^{n_{u}}$, to form a larger tensor $\bm{X}$ with dimension $(c+n_{u}) \times H/2\times W/2$. Note that $\bm{u}$ is obtained by feeding both $\mathrm{SNR}$ and $\rho_l$ ($\rho_l$ is not needed if the successive refinement scheme is adopted) to a fully connected layer and $\bm{X}$ can be obtained as:
\begin{align}
       &\bm{u} = \mathrm{MLP}([\mathrm{SNR}, \rho_l]) \notag \\
    \bm{X} = \widetilde{\bm{X}} & \oplus \bm{U}_e  \quad \text{with} \; \bm{U}_e[:,i,j] =  \bm{u},
\end{align}
where $\bm{U}_e\in \mathbb{R}^{n_u\times H/2\times W/2}$ is obtained by duplicating the vector $\bm{u}$ and $\oplus$ denotes concatenation. Then $M_1$ Swin transformer blocks\footnote{{Note that $M_1$ is required to be an even number since the shifted window self-attention is performed over two consecutive transformer blocks.}} are applied to $\bm{X}$ before proceeding into the next stage. We briefly introduce the operations of the Swin transformer block and refer readers to \cite{swin} for more details.

The Swin transformer blocks resemble those in the standard vision transformer \cite{vit} except the multi-head self-attention module is replaced by the one based on shifted windows. We describe the operations of two consecutive Swin transformer blocks as follows. Assume that the input feature tensor $\bm{X}_i$ at stage $i$ has dimension $c\times h_i\times w_i$ and each window contains $w\times w$ patches/tokens, then the first Swin transformer block whose operation is denoted by $\textbf{W-MSA}$, evenly partitions $\bm{X}_i$ into $(h_i/w, w_i/w)$ non-overlapping windows\footnote{We assume that $h_i$ and $w_i$ are multiples of $w$. We can simply zero-pad the corresponding tensors if the assumption does not hold.} then performs multi-head self-attention within each window. To allow communications between different windows, a shifted window multi-head self-attention operation, denoted as $\textbf{SW-MSA}$, is adopted, where a cyclic shift is applied to the window configuration in the first transformer block followed by the multi-head self-attention operated within the new (shifted) windows. We summarize the operations as follows:
\begin{align}
    \hat{\bm{X}}_i^{(1)} &= \textbf{W-MSA}(\text{LN}(\bm{X}_i)) + \bm{X}_i \notag \\
    {\bm{X}}_i^{(1)} &= \text{MLP}(\text{LN}(\hat{\bm{X}}_i^{(1)})) + \hat{\bm{X}}_i^{(1)} \notag \\
    \hat{\bm{X}}_i^{(2)} &= \textbf{SW-MSA}(\text{LN}(\bm{X}_i^{(1)})) + \bm{X}_i^{(1)} \notag \\
    {\bm{X}}_i^{(2)} &= \text{MLP}(\text{LN}(\hat{\bm{X}}_i^{(2)})) + \hat{\bm{X}}_i^{(2)}.
    \label{eq:swin_trans}
\end{align}
where $\text{LN}$ denotes the layer normalization and we use the superscripts to distinguish the features processed by the first and second Swin transformer block. By applying the operations defined in \eqref{eq:swin_trans} $M_i/2$ times, we obtain the final output of the $i$-th stage.

As shown in Fig. \ref{fig:detailed_net}, the output of the Swin transformer blocks in the $i$-th stage is then fed to the patch merging module in the $(i+1)$-th stage, whose output is of dimension $c \times h_{i+1} \times w_{i+1}$. Note that, in our setting, the patch merging layer concatenates the features of $2 \times 2$ neighboring patches and applies a linear layer to reduce the $4c$-dimensional features to a dimension of $c$. Thus, we have $(h_{i+1}, w_{i+1}) = (h_i/2, w_i/2)$. After passing all the $I$ stages, we obtain the output with dimension $c\times h_{I}\times w_{I}$, which is reshaped and linear projected to matrix $\widetilde{\bm{Z}}$ with dimension $N_F \times N_T$, where $N_F N_T = 2\rho_L N$ with $N_T = h_I w_I$ .

\textbf{Varying Patches versus Varying Features.} The matrix $\widetilde{\bm{Z}}$ contains $N_T$ tokens, each consisting of $N_F$ features for the maximum bandwidth ratio $\rho_L$. To be adaptive to different $\rho_l$'s, one may either transmit a reduced number of tokens $n_t < N_T$, while keeping the dimension of features per token the same ($n_f = N_F$), which is called \textit{varying patches}, or reduces the dimension of features per token (we assume the same $n_f$ for different tokens) while fixing the number of tokens ($n_t = N_T$), which is referred to as \textit{varying features}. Note that a more flexible bandwidth adaptive scheme is proposed in \cite{bupt}, where different $n_f$'s are assigned to different tokens for better performance. However, we argue that this requires transmitting additional digital information to inform the decoder concerning how to partition the received signal for each token. When the digital information is not correctly decoded, the decoder would totally fail to reconstruct the original image. Our schemes, both the \textit{varying patches} and the \textit{varying features}, on the other hand, do not require to transmit any digital information\footnote{We assume that $\rho_l$ is available to both the transmitter and receiver as part of the control channel information.}. We evaluate the reconstruction performance for the two schemes, and find that they yield similar performance, which will be detailed in Section \ref{sec:IV}. In the following discussions, we employ the \textit{varying features} scheme.

\textbf{Data flow at the decoder.} Upon receiving $\bm{y}_l \in \mathbb{C}^{\rho_lN}$, the receiver first converts it to a real tensor and then zero-pads it along the first dimension to obtain $\widetilde{\bm{Y}}_l \in \mathbb{R}^{N_F \times h_I \times w_I}$, denoted by $\widetilde{\bm{Y}}_l = \mathcal{P}(\bm{y}_l)$, where $\mathcal{P}$ represents the reshaping and zero-padding operations. As shown in Fig. \ref{fig:detailed_net}, the same side information $\bm{u}$ for the encoder is first duplicated to $\bm{U}_{d}$ and then concatenated to each token of $\widetilde{\bm{Y}}_l$ to form ${\bm{Y}}_l \in \mathbb{R}^{(N_F+n_{u}) \times h_I \times w_I}$ whose $h_I \times w_I$ tokens will be further mapped to a $c$-dimensional vector. The decoder also has $I$ stages and each stage consists of a patch division block and Swin transformer blocks. To be precise, the patch division block adopts pixel shuffling to upsample the spatial dimension of the input tensors. The upsampled tensor is fed to the Swin transformer blocks in the subsequent stage, whose structures are identical to those at the encoder. After passing all $I$ stages, a 2d-convolutional layer converts the latent tensor to the reconstructed image $\widetilde{\bm{S}}_l \in \mathbb{R}^{C\times H \times W}$.

\begin{algorithm}[t!]
\caption{Overall Training Process for DeepJSCC-l++ Model with DWA.}\label{alg:train_process}
\begin{algorithmic}[1]
\State{\textbf{Initialize} $w_l^1 = 1, \forall l \in [L]$}
\For{$t=1,\ldots,T$}
\State{\textbf{Training Phase}:}
\For{each batch}
\State{\textbf{Sample} $l \in [L], \mathrm{SNR} \in [\mathrm{SNR}_{min}, \mathrm{SNR}_{max}]$}
\State{\textbf{Encoder:} $\bm{z}_l = f_{\Theta}(\bm{S}, \mathrm{SNR}, l)$}
\State{\textbf{Decoder:} $\widetilde{\bm{S}}_l = g_{\Psi}(\bm{y}, \mathrm{SNR}, l)$}
\State{\textbf{Weighted Loss:} $\mathcal{L}_l^t = w_l^t||\bm{S} - \widetilde{\bm{S}}_l||^2_2$.}
\State{\textbf{Optimize} $\{\Theta, \Psi\}$ using $\mathcal{L}_l^t$}.

\EndFor
\State{\textbf{Validation Phase}:}
\For{$l \in [L]$} \State{\textbf{Calculate} $\text{PSNR}^t_l, \Delta_l^t$ over validation set.}

\State{\textbf{Update} $w_l^t$ according to \eqref{eq:policy}. }
\EndFor
\EndFor
\end{algorithmic}
\end{algorithm}

\subsection{Dynamic Weight Assignment (DWA)}\label{sec:IIIB}
In this subsection, we introduce a training methodology designed for both the successive refinement and adaptive-bandwidth frameworks, which is essential for good reconstruction performance with different bandwidth ratios. 

To train a single model that is adaptive to the bandwidth ratio $\rho_l$ and the channel SNR, we uniformly sample $\rho_l \in \{\rho_1, \ldots, \rho_L\}$ and $\mathrm{SNR} \in [\mathrm{SNR}_{min}, \mathrm{SNR}_{max}]$ and feed them along with the image $\bm{S}$ to both the encoder and decoder during training. The loss $\mathcal{L}_{l}$ for $\rho_l$ is evaluated using the mean square error (MSE) between the input image and its reconstruction.
For small $\rho_l$, $\mathcal{L}_{l}$ is much larger compared with that of larger $\rho_l$. As the losses for all bandwidth ratios are optimized together, the proposed DeepJSCC-l++ tends to focus on the reconstruction performance for smaller bandwidth ratios with significantly larger loss values. The reconstruction for larger bandwidth ratios, however, becomes highly sub-optimal, which motivates us to explore a better training methodology to improve the reconstruction quality across all conditions. 

\begin{figure}[t]
  \centering
  \includegraphics[width=0.65\linewidth]{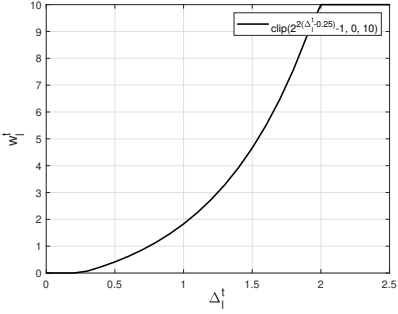}\\
  \caption{We plot the function in \eqref{eq:policy} for the parameters $\alpha = 2, \beta = 0.25, \Gamma = 10$.}
\label{fig:function}
\end{figure}

Inspired by \cite{elf_vc}, we propose the DWA scheme, which assigns different weights $w_l^t$ to the loss achieved for different $\rho_l$ values according to their reconstruction qualities in the validation phase at the $t$-th epoch. To be specific, we first obtain an average image reconstruction quality $\text{PSNR}_l^t$ as defined in \eqref{eq:psnr} for each bandwidth ratio $\rho_l$ in the $t$-th epoch, where the images from the validation dataset along with $\rho_l$ and $\mathrm{SNR}_{val}$ are fed to the DeepJSCC-l++ model. {Without loss of generality, we set the $\mathrm{SNR}_{val} = (\mathrm{SNR}_{min}+\mathrm{SNR}_{max})/2$}. To evaluate the relative reconstruction performance for each $\rho_l$, we compare $\text{PSNR}_l^t$ with the PNSR \textit{upper bound} for that specific bandwidth ratio $\rho_l$, denoted by $\text{PSNR}_l^*$, which is obtained by training a non-adaptive model with fixed $\mathrm{SNR}_{val}$ and $\rho_l$. Intuitively, a larger gap from $\text{PSNR}_l^*$, requires a larger weight $w_l^t$ assigned to that bandwidth ratio $\rho_l$. Based on this intuition, we introduce a policy to dynamically assign weights to the losses, $\mathcal{L}_{l}^t$, for different bandwidth ratios:
\begin{align}
    &\Delta_l^t = \text{PSNR}_l^* - \text{PNSR}_l^t, \notag \\
    w_l^t &= \text{clip}(2^{\alpha (\Delta_l^t - \beta)}\! - \! 1, \: 0, \Gamma),\label{eq:policy}
\end{align}
where $\alpha$ is set to be a positive number to ensure that a larger weight, $w_l^t$, is assigned to the bandwidth ratio $\rho_l$ with a larger $\Delta_l^t$.  Since the reconstruction quality of the adaptive model is expected to be \textit{upper bounded} by separately (non-adaptive) trained models, we would expect a PSNR gap of the adaptive model from these benchmarks. Thus, we set a non-zero $\beta$, which allows the proposed DeepJSCC-l++ model to focus on optimizing the reconstruction performance for other bandwidth ratios $\rho_i, i\neq l$ if $\Delta_l^t$ is smaller than $\beta$. Note that, even when $\Delta_l^t < \beta$, a non-negative weight is required. Thus, we clip the weights to have a minimum value $0$ and a maximum value $\Gamma$ to ensure stable training. Empirically, we found that setting $\alpha = 2, \beta = 0.25, \Gamma = 10$ yields reasonable performance and we use these values throughout the paper. Fig. \ref{fig:function} plots the function in \eqref{eq:policy} for the settings given above. {We can observe the weight assigned to a certain bandwidth ratio decays smoothly with the reduction of $\Delta_l^t$ during the training process.} The overall training process with the DWA training methodology for DeepJSCC-l++ is summarized in Algorithm \ref{alg:train_process}.

\section{Numerical Experiments}\label{sec:IV}

Next, we evaluate the performance of the proposed DeepJSCC-l++ scheme for both the adaptive-bandwidth and successive refinement scenarios, along with the non-adaptive models trained at a fixed bandwidth ratio and channel SNR\footnote{Code available at \url{https://github.com/aprilbian/deepjscc-lplusplus}.}.

\subsection{Parameter Setting and Training Details}
We consider the transmission of images from the CIFAR-10 dataset, which includes 40000 training, 10000 validation, and 10000 test RGB images, each with $32\times 32$ resolution. Both the encoder and decoder employ $I = 2$ stages, the number of features $c$ is set to 256, the window size to $w = 8$, the numbers of Swin transformer blocks in each stage to $M_1 = 4, M_2 = 2$. The dimension of the embedding is set to $n_{u} = 2$. Finally, the GeLU activation function is used within the Swin transformer blocks. 

For training, we adopt the Adam optimizer and a varying learning rate initialized at $10^{-4}$, which is reduced by a factor of $0.95$ if the validation loss does not drop for 20 epochs. The maximum number of epochs is set to $4\times 10^3$, and the early stopping module is included, where the training process terminates if the validation loss does not improve in $80$ epochs for the adaptive schemes, whereas the patience is set to $60$ for the non-adaptive models. Note that the settings mentioned above ensure that the validation losses are saturated for both adaptive and non-adaptive schemes when the training ends.

Throughout this section, we assume a maximum bandwidth ratio of $\rho_{max} = 1/4$, which corresponds to a maximum number of tokens $N_T = 64$ and maximum number of features per token $N_F = 24$. We train and evaluate the proposed schemes with channel SNR ranging from $4$ dB to $10$ dB.

\subsection{Performance Evaluation}
We first compare the reconstruction performance of the bandwidth-adaptive model with \textit{varying patches} and \textit{varying features}. In this experiment, we set the number of supported bandwidth ratios to $L = 4$ with $\rho_l \in \{1/16, 1/8, 3/16, 1/4\}$. The \textit{varying patches} scheme adopts a fixed $N_F = 24$ while its $n_t$ changes from $16$ to $64$ to adjust for different bandwidth ratios, the \textit{varying features} scheme, on the other hand, fixes $N_T = 64$ with a varying number of features $n_f = \{6, 12, 18, 24\}$. The PSNR performance of the two adaptive schemes as well as the separately trained  (non-adaptive) models under a fixed channel quality $\mathrm{SNR} = 7$ dB are shown in Table I. Note that both adaptive schemes adopt the DWA training methodology proposed in Section \ref{sec:IIIB}. We observe similar PSNR values for the two schemes, and both of them achieve comparable performance with the separately trained models where a maximum of $0.24$ dB PSNR gap is observed at $\rho_L = 1/4$. We use the \textit{varying features} scheme for the remaining simulations.

\begin{table}[tbp]
\caption{{Evaluation for the \textit{varying patches} and \textit{varying features} DeepJSCC-l++ schemes at $\mathrm{SNR} = 7$ $dB$ in terms of PSNR ($dB$).}}
\begin{center}
\begin{tabular}{c|cccc}
\hline

\cline{1-5} 
\textbf{$\rho$} & \textbf{1/16}& \textbf{1/8}& \textbf{3/16} & \textbf{1/4}\\
\hline

\textit{varying patches} & 26.12  & 30.01& 32.53& 34.32\\
\textit{varying features} & 26.14 & 30.01 & 32.53& 34.31\\
\textit{separate training} & 26.36& 30.23& 32.70& 34.55\\

\hline
\end{tabular}
\label{tab:patch_or_feature}
\end{center}

\end{table}

\begin{figure*}
     \centering
     \begin{subfigure}{0.65\columnwidth}
         \centering
         \includegraphics[width=\columnwidth]{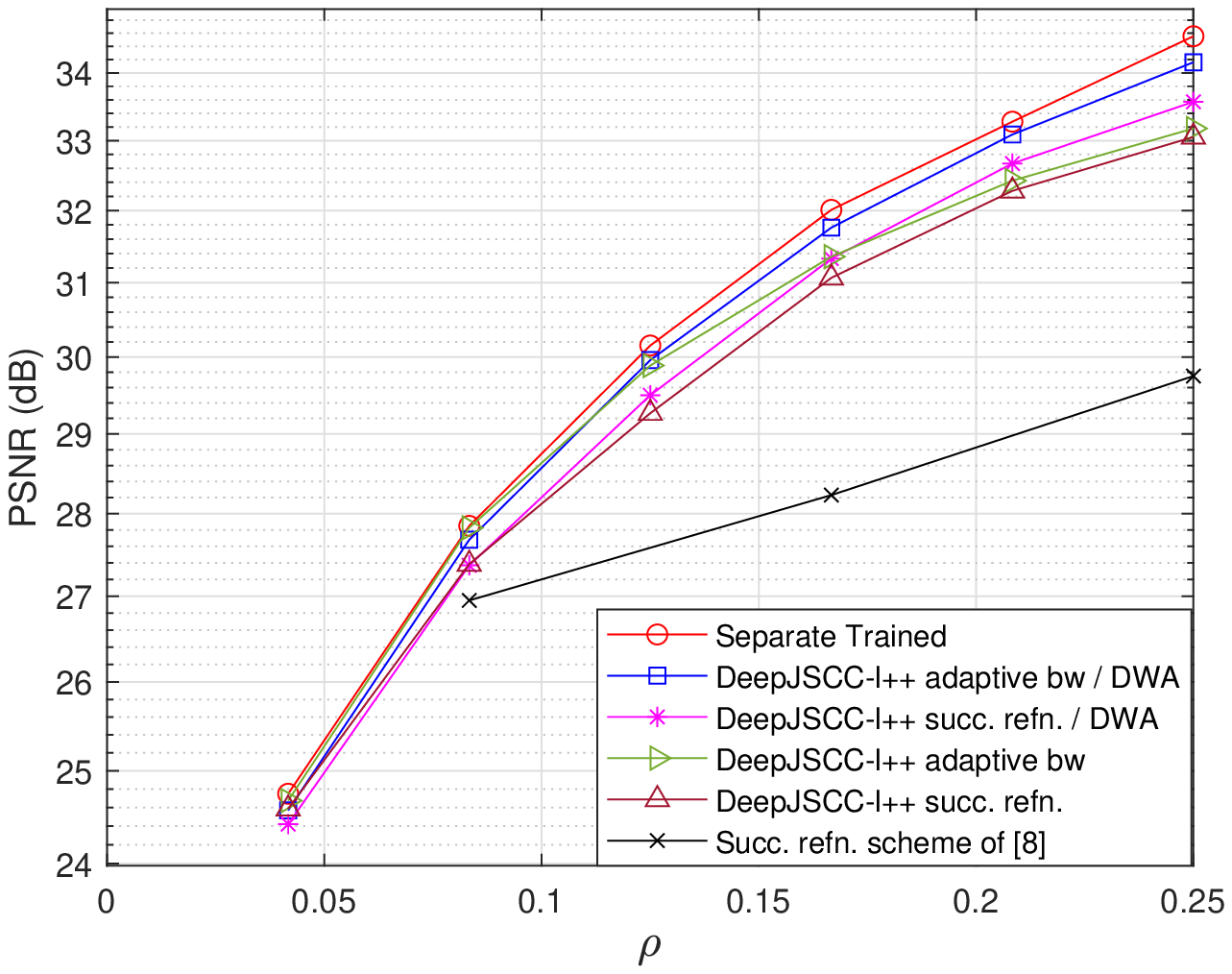}
     \end{subfigure}
     \begin{subfigure}{0.65\columnwidth}
         \centering
         \includegraphics[width=\columnwidth]{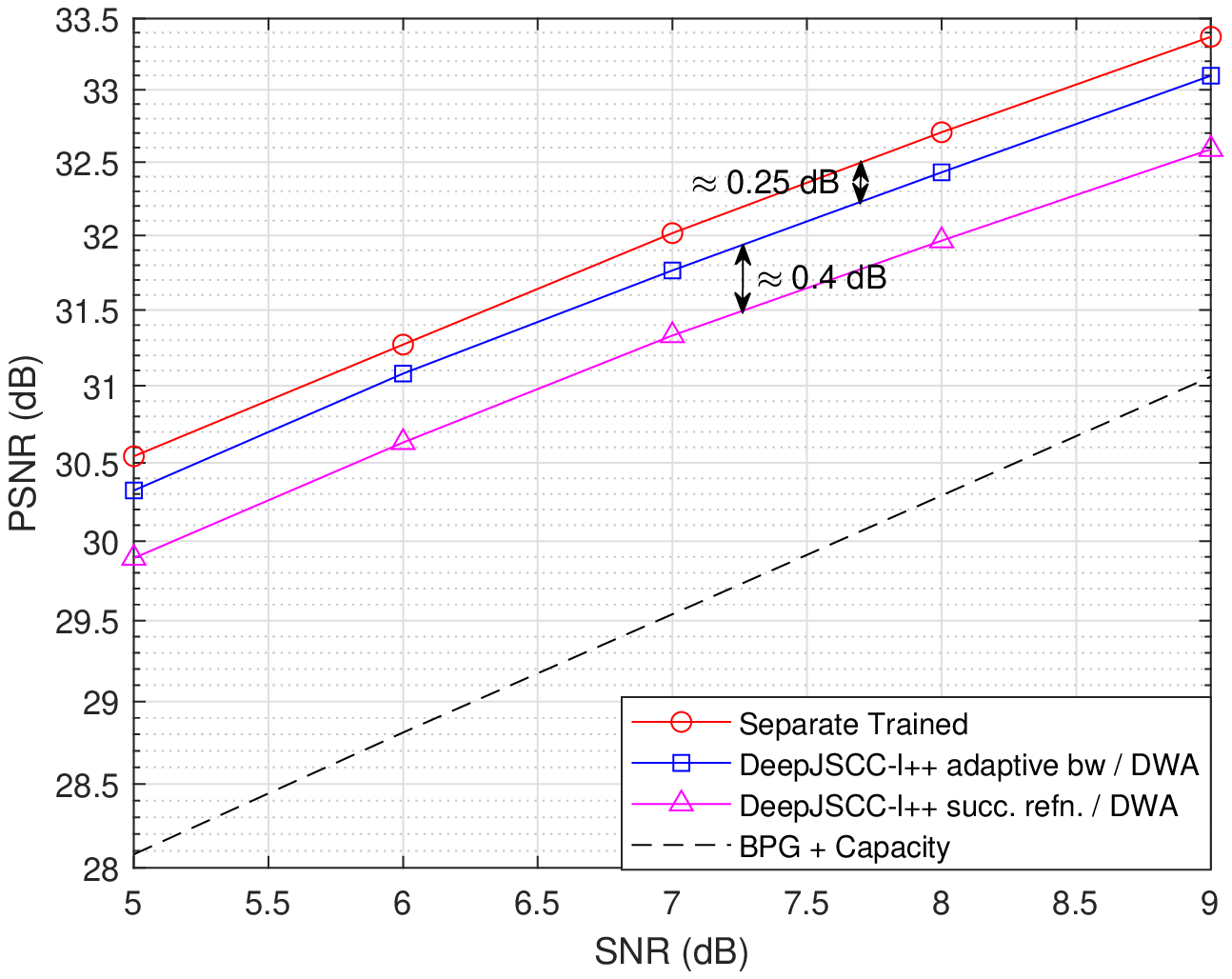}
     \end{subfigure}
     \begin{subfigure}{0.65\columnwidth}
         \centering
         \includegraphics[width=\columnwidth]{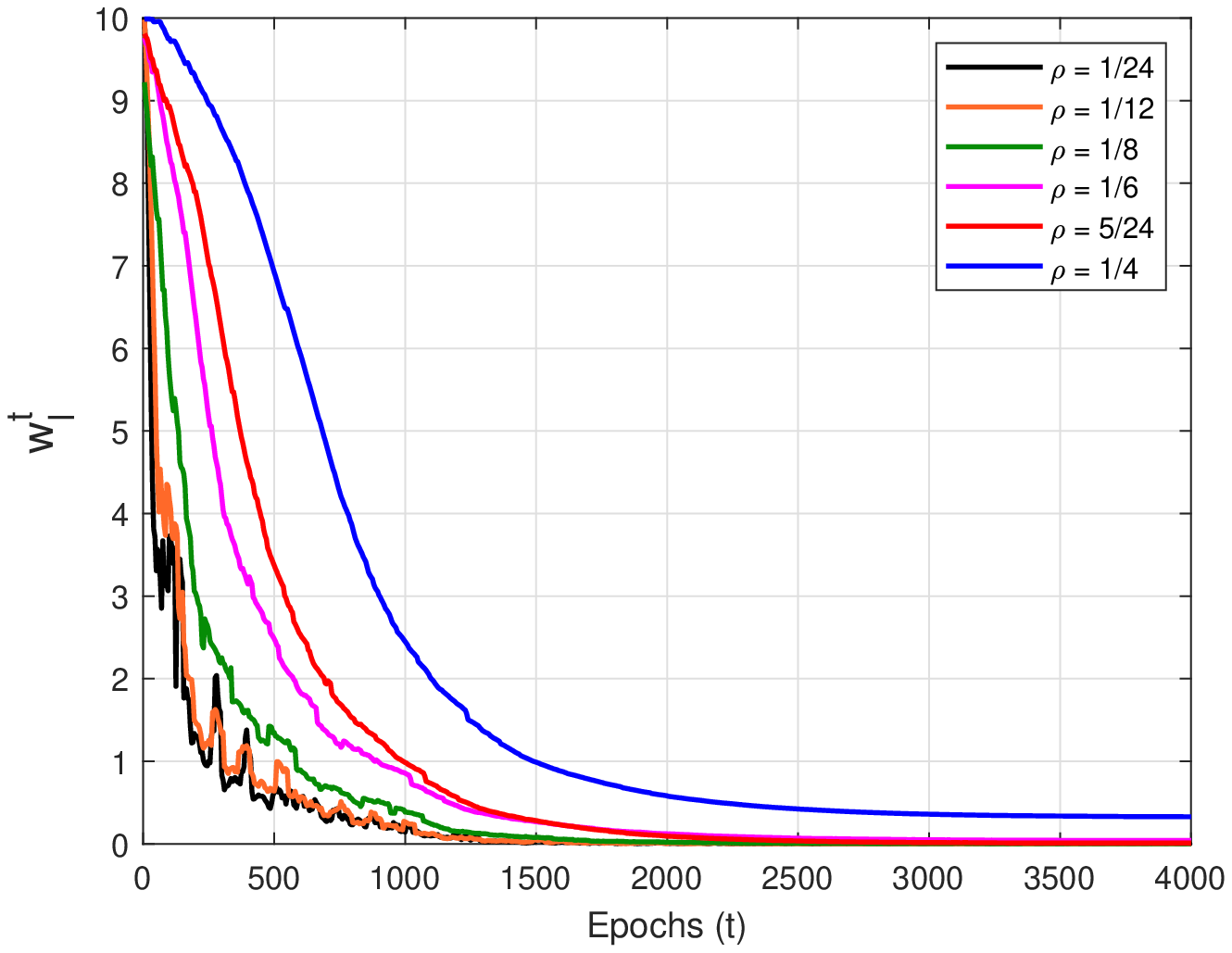}
     \end{subfigure}
  \caption{We show the effectiveness of the proposed DeepJSCC-l++ models: (a) the PSNR performance over different $\rho_l$ with $\mathrm{SNR} = 7$ dB;  (b) the PSNR performance over different evaluation SNRs with $\rho_l = 1/6$; (c) the weights $w_l^t$ for different bandwidth ratios $\rho_l$ with respect to the number of epochs $t$. }
\label{fig:final_simu}
\end{figure*}

Next we consider a more challenging case where we train the adaptive models with $L = 6$ supported bandwidth ratios and varying channel SNRs. The possible bandwidth ratios are $\rho_l \in \{1/24, 1/12, 1/8, 1/6, 5/24, 1/4\}$, which correspond to $n_f = \{4, 8, 12, 16, 20, 24\}$. We show the PSNR performance of the {proposed DeepJSCC-l++ applied to both bandwidth-adaptive and successive refinement scenarios} introduced in Section \ref{sec:II} (with and without DWA), along with the DeepJSCC-l model originally proposed in \cite{deepjsccl}, which follows the successive refinement principle but adopts a CNN as its backbone.

As can be seen in Fig. 4 (a) and (b), where we fix the channel SNR to $7$ dB, the proposed DeepJSCC-l++ model with DWA can be adaptive to different bandwidth ratios with a negligible performance gap from the separately trained (non-adaptive) benchmarks. Specifically, we observe that for the proposed DeepJSCC-l++ models, the gap is more significant at larger $\rho_l$ values, {which is intuitive as the optimization task for reconstructing the image from a higher dimensional latent is in general more challenging}\footnote{This can also be verified by checking the training curves for different $\rho_l$ values.}. Further, it is shown that the {DeepJSCC-l++ under successive refinement scenario} with DWA generates a reconstruction performance with a modest PSNR gap of $\approx 0.5$ dB compared to the adaptive-bandwidth scenarios at high $\rho_l$ values. This is due to the fact that the successive refinement scheme imposes more stringent constraints as illustrated in Section \ref{sec:II}. The DeepJSCC-l++ without the proposed training methodology, on the other hand, is optimized with overwhelming focus on the performance of the smallest bandwidth ratio, $\rho_1 = 1/24$, yielding $\Delta_1^T \approx 0$ dB where $T$ denotes the epoch where the training terminates. However, the reconstruction performances at higher $\rho_l, l\!>\!1$, are far from optimal, which highlights the effectiveness of the proposed DWA training methodology. The curve named `{Succ. refn. scheme of \cite{deepjsccl}}' in Fig. \ref{fig:final_simu} (a) represents the results obtained in \cite{deepjsccl} for successive refinement, which has fallen short of the performance of the DeepJSCC-l++ scheme which is mainly due to the less powerful CNN backbone adopted in \cite{deepjsccl}. 

We then demonstrate that the proposed DeepJSCC-l++ model is also adaptive to different channel SNRs. In this simulation, the settings are the same as those in Fig. \ref{fig:final_simu} (a). The bandwidth ratio $\rho_l$ is fixed at $1/6$ while the evaluation $\mathrm{SNR}$ varies from $5$ dB to $9$ dB. As shown in Fig. \ref{fig:final_simu} (b), the PSNR performance of the DeepJSCC-l++ models are compared with the separately trained models. A fixed $\approx 0.25$ dB PSNR gap from the separately trained models is observed for the DeepJSCC-l++ model, which corresponds to the $\Delta_l^T$ value introduced in Section \ref{sec:IIIB} and also shown in Fig. \ref{fig:final_simu} (a). All these schemes show significant gain compared with the digital baseline which utilizes the BPG image compression and capacity achieving code. Combining the results in Fig. \ref{fig:final_simu} (a) and (b), we confirm the proposed schemes are adaptive to different bandwidth ratios and channel SNRs.

Finally, for a comprehensive understanding of the DWA training methodology for the DeepJSCC-l++, we present the weights $w_l^t$ versus the number of epochs $t$ for different bandwidth ratios $\rho_l$'s in Fig. \ref{fig:final_simu} (c). Note that the settings for this simulation is identical to that in Fig. \ref{fig:final_simu} (a). As analyzed before, {the reconstruction task for larger $\rho_l$ is generally harder}, thus for a fixed epoch $t$, the corresponding $\Delta_l^t$ is larger leading to a larger $w_l^t$ which is verified in the figure. As the reconstruction performance improves with more epochs, some of the $w_l^t$'s reduce to 0 if $\Delta_l^t \le \beta = 0.25$. Since a PSNR gap greater than $0.25$ dB remains when the training finishes, we can still observe a non-zero weight for the $\rho_L = 1/4$ case.

\section{Conclusion}\label{sec:6}
We developed a flexible wireless image transmission scheme, called DeepJSCC-l++, which can adapt to multiple bandwidth ratios and channel SNRs using a single encoder/decoder neural network pair built upon the Swin transformer architecture. 
Numerical simulations verify that the proposed DeepJSCC-l++ model can be adaptive to both the bandwidth ratio and the channel SNR with marginal performance loss with respect to the separately trained models. It is also shown that the proposed architecture can be applied to the successive refinement problem, outperforming the state-of-the-art by a significant margin. It also outperforms the separation-based baseline considering BPG compression with a capacity-achieving channel code, which provides an upper bound on the performance achievable by a separation-based scheme employing BPG for compression. These results not only show the superior performance of DeepJSCC, but also highlight its potential for practical systems through the use of a single pair of encoder/decoder parameters. 

\appendices


\bibliographystyle{IEEEtran}
\bibliography{References}

\end{document}